\documentclass[aps,reprint,prl,twocolumn,showpacs,amsmath,longbibliography]{revtex4}  
\usepackage{bbm}
\usepackage{mathrsfs}
\usepackage{amsmath}
\usepackage{amsfonts}
\usepackage[colorlinks=true,citecolor=blue,anchorcolor=blue]{hyperref}
\usepackage{graphicx,epstopdf}
\usepackage{subfigure}
\usepackage{epsfig}
\usepackage{dcolumn}
\usepackage{bm}
\usepackage{color}
\usepackage{natbib}
\usepackage{amssymb}
\usepackage{xcolor}
\usepackage{braket}

\begin{document}

\title{Three-Dirac-fermion approach to unexpected universal gapless surface states in van der Waals magnetic topological insulators}

\author{Dinghui Wang$^{1}$, Huaiqiang Wang$^{1,\ast}$, Dingyu Xing$^{1,2}$, and Haijun Zhang$^{1,2,\ast}$}

\affiliation{
 $^1$ National Laboratory of Solid State Microstructures, School of Physics, Nanjing University, Nanjing 210093, China\\
$^2$ Collaborative Innovation Center of Advanced Microstructures, Nanjing University, Nanjing 210093, China\\
}

\begin{abstract}
Layered van der Waals (vdW) topological materials, especially the recently discovered MnBi$_2$Te$_4$-family magnetic topological insulators (TIs), have aroused great attention. However, there has been a serious debate about whether the surface states are gapped or gapless for antiferromagnetic (AFM) TI MnBi$_2$Te$_4$, which is crucial to the prospect of various magnetic topological phenomena. Here, a minimal three-Dirac-fermion approach is developed to generally describe topological surface states of nonmagnetic/magnetic vdW TIs under the modulation of the interlayer vdW gap. In particular, this approach is applied to address the controversial issues concerning the surface states of vdW AFM TIs. Remarkably, topologically protected gapless Dirac-cone surface states are found to arise due to a small expansion of the interlayer vdW gap on the surface, when the Chern number equals zero for the surface ferromagnetic layer; while the surface states remain gapped in all other cases. These results are further confirmed by our first-principles calculations on AFM TI MnBi$_2$Te$_4$. The theorectically discovered gapless Dirac-cone states provide a unique mechanism for understanding the puzzle of the experimentally observed gapless surface states in MnBi$_2$Te$_4$. This work also provides a promising way for experiments to realize the intrinsic magnetic quantum anomalous Hall effect in MnBi$_2$Te$_4$ films with a large energy gap.
\end{abstract}


\email{zhanghj@nju.edu.cn;\\hqwang@nju.edu.cn}

\maketitle

\section{I. Introduction} 
Three-dimensional topological insulators (TIs) are characterized by gapless Dirac-cone surface states in the bulk band gap~\cite{Hasan2010rmp, Qi2011rmp}. However, breaking the time-reversal symmetry (TRS) leads to an energy gap at the Dirac point of these surface states~\cite{Chen2010massive}. Such gapped Dirac-cone surface states play a crucial role in a series of exotic topological quantum phenomena~\cite{Tokura2019magnetic}, including the topological magnetoelectric effect~\cite{Qi2008prb,Wang2015prb} and the quantum anomalous Hall effect (QAHE)~\cite{Yu2010quantized, Chang2013experimental}. While magnetic doping has conventionally been used to induce gapped Dirac-cone surface states in TIs~\cite{Chang2013experimental}, this method is inevitably subject to complex and detrimental effects, such as magnetic inhomogeneity and disorder effects. Recently, MnBi$_2$Te$_4$-family intrinsic magnetic TIs were discovered~\cite{Gong2019cpl, Otrokov2019nature, Zhang2019mbt, Li2019sa, Chen2019intrinsic, Otrokov2019prl, Vidal2019surface, Liu2020robust, Ge2020high, Yan2019prm, Yan2021prb, Sun2019rational, Gu2021spectral, Sun2020prb, Li2021prl, Li2021critical, Gao2021layer, Fu2020exchange, Vidal2019topological, Hu2020van,Lian2020prl, Zeugner2019chemical, Wu2019natural,Klimovskikh2020tunable,Shi2019prb,Sass2020prl,Yuan2020electronic, Li2020antiferromagnetic,Zhang2019experimental,Ying2022prb,xie2020mechanism,Li2020intrinsic,zhong2021light,liu2021magnetic,Xu2022prb, Chen2021Koopman, garnica2022native}, which offers a promising alternative playground. These materials not only increase the temperature of the QAHE~\cite{Deng2020quantum}, but also enable the study of axion electrodynamics~\cite{Zhang2020Mn2Bi2Te5, Wang2020heterostructures, Zhu2021tunable, Liu2020prb,Xiao2021prb, Sekine2021axion}. However, there is still a significant controversy on the surface states of MnBi$_2$Te$_4$, as both gapped and gapless surface states were experimentally observed~\cite{Otrokov2019nature, Shikin2021prb, Chen2019prx, Li2019prx, Hao2019prx, Ji2021detection, Lee2019prr, Nevola2020prl}. 

\begin{figure}[htbp]
\includegraphics[width=3in]{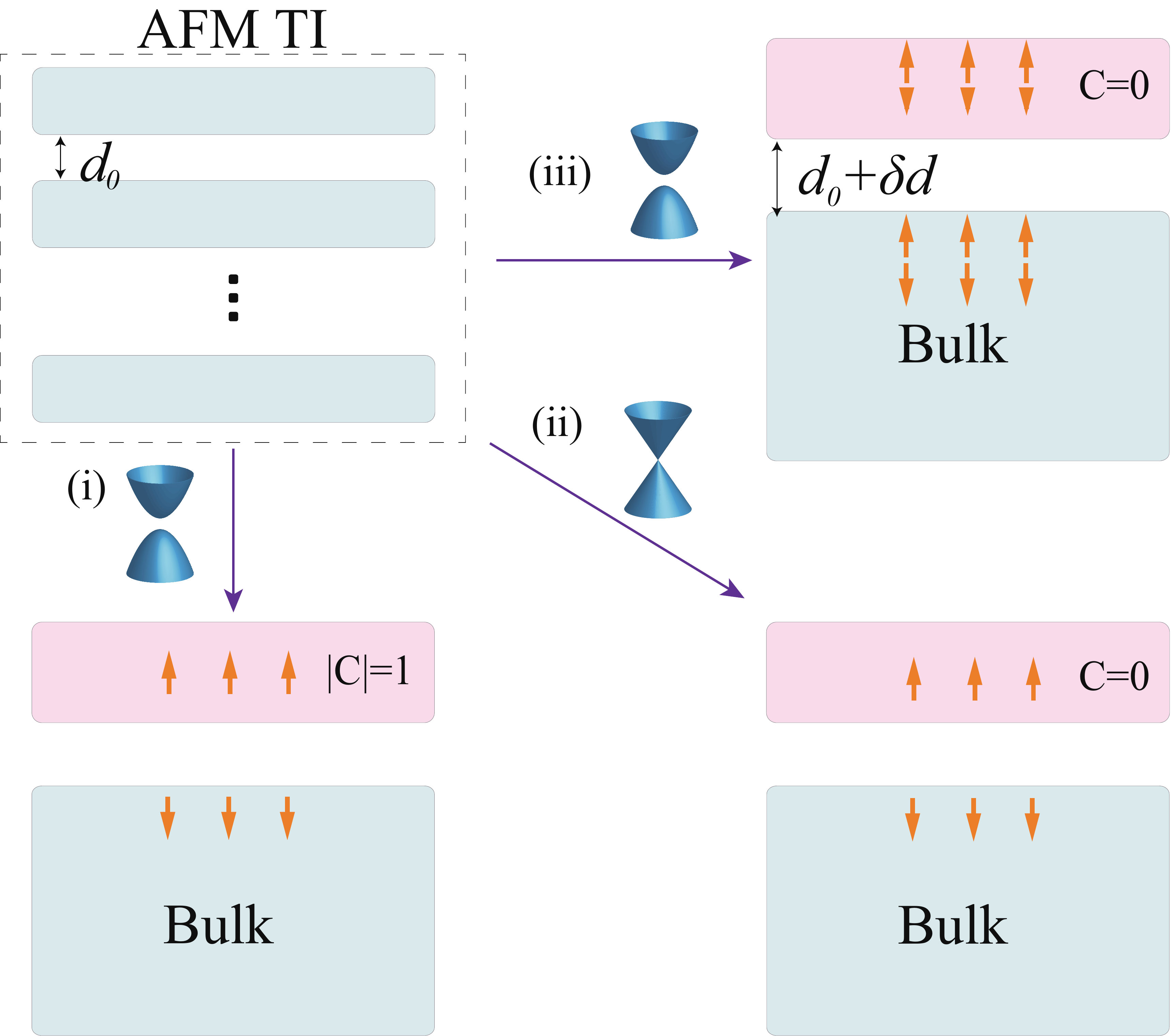}
\caption{Evolution of surface states of an AFM TI with an expansion of the topmost vdW gap. The AFM TI consists of FM or AFM block layers (colored in blue) held together by weak vdW forces. In the unexpanded case ($d_0$), the AFM TI has gapped Dirac-cone surface states. With increasing the topmost vdW gap ($d=d_0+\delta d$) between a surface layer (colored in pink) and the below bulk (colored in blue), if the surface layer is FM, two cases will occur, depending on its Chern number $C$: (i) When $|C|=1$, the surface states remain gapped. (ii) When $C=0$, the energy gap of surface states has a topologically protected gapless transition point. In contrast, (iii) if the surface layer is AFM, the surface states always keep gapped.}
\label{fig1}
\end{figure}

Meanwhile, most nonmagnetic/magnetic TIs are layered materials with an interlayer van der Waals (vdW) gap, such as Bi$_2$Se$_3$-family TIs \cite{Zhang2009topological} and MnBi$_2$Te$_4$-family magnetic TIs \cite{Gong2019cpl, Otrokov2019nature, Zhang2019mbt, Li2019sa}. These materials are composed of covalent-bonding layers held together by weak vdW forces. The interlayer vdW gap plays a crucial role in the electronic structures of both bulk and surface states of these materials. However, the vdW gap is sensitive to impurities or intercalated atoms in fabrication processes, and even a small concentration of impurities can significantly expand the vdW gap.  Moreover, the surface-induced symmetry breaking may also cause a considerable expansion of the topmost vdW gap. Previous studies have shown that an expansion of the topmost vdW gap can lead to a relocation of surface states in Bi$_2$Se$_3$-family TIs~\cite{Eremeev2012effect}.

In this work, we develop a three-Dirac-fermion model that can generally describe the surface states of layered nonmagnetic/magnetic TIs under the modulation of the interlayer vdW gap. We apply this model to investigate the evolution of the surface states of layered A-type AFM TIs~\cite{Mong2010prb}, as the topmost vdW gap expands from the bulk value ($d=d_0$) to the surface layer decoupled limit ($d=\infty$), as illustrated in Fig.~\ref{fig1}. The surface layer can be regarded as either an effective ferromagnetic (FM) block consisting of an odd number of septuple-layers (SLs) of MnBi$_2$Te$_4$~\cite{Zhang2019mbt}, as schematically shown in Fig.~\ref{fig1}(i,ii), or an effective AFM block consisting of an even number of SLs of MnBi$_2$Te$_4$, as schematically shown in Fig.~\ref{fig1}(iii). For the FM surface layer, we find that if the surface layer has a zero Chern number, unexpected topologically protected gapless Dirac-cone surface states will arise at the topological transition point [see (ii) in Fig.~\ref{fig1}], while the gapped surface states persist if the surface layer has a nonzero Chern number [see (i) in Fig.~\ref{fig1}]. On the other hand, for the AFM surface layer, the energy gap of surface states is always maintained [see (iii) in Fig.~\ref{fig1}]. These findings are further confirmed by our first-principles calculations on AFM TI MnBi$_2$Te$_4$, which may solve the puzzle of the angle resolved photoelectron spectroscopy (ARPES) observations of both gapless and gapped surface states of MnBi$_2$Te$_4$. Notably, the three-Dirac-fermion approach provides a unified description of surface states of layered TIs, indicating that engineering the interlayer vdW gap can provide a new route for developing the applications of nonmagnetic/magnetic TIs.

\begin{figure*}[htbp]
\includegraphics[width=5.8in]{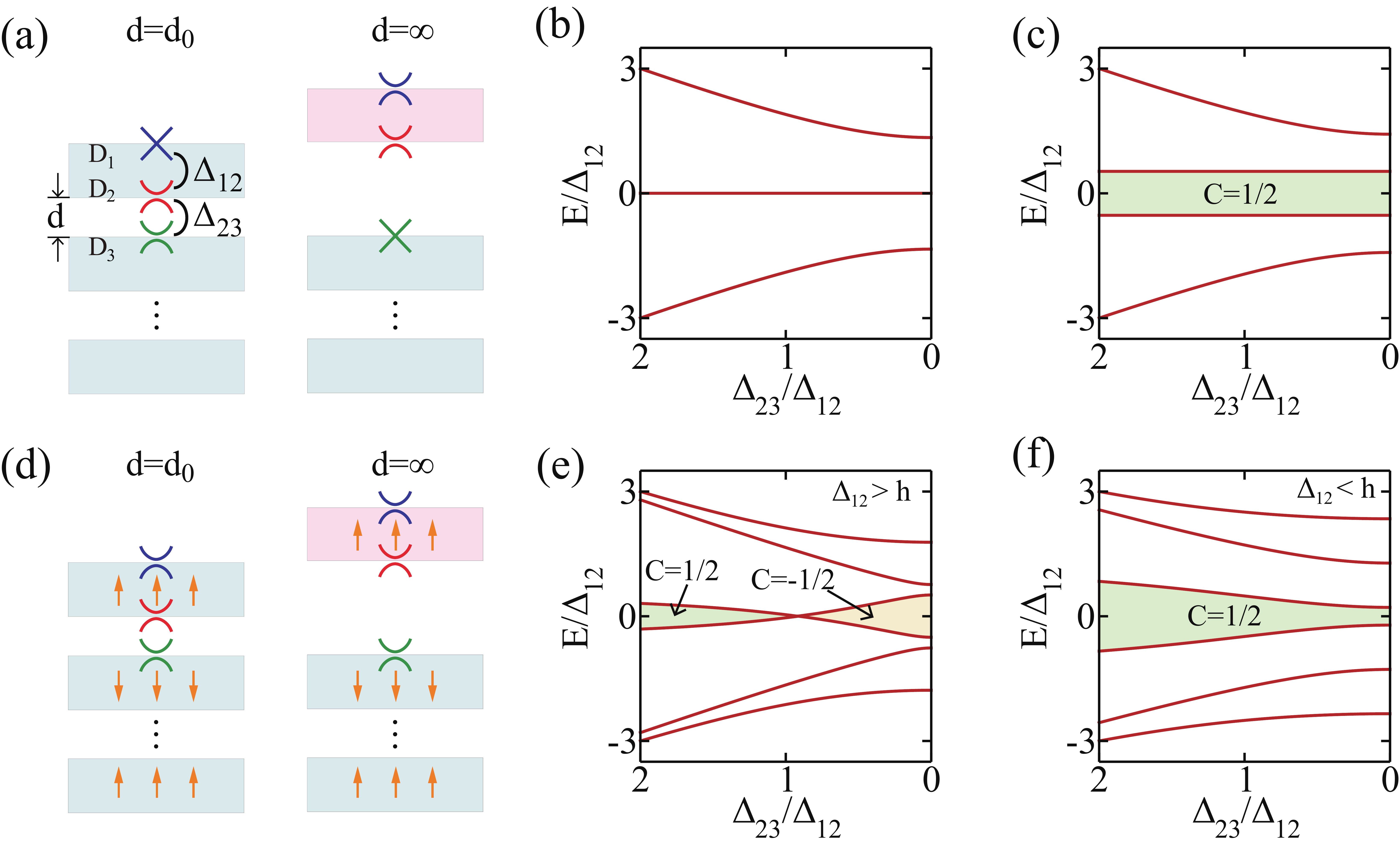}
\caption{The three-Dirac-fermion model. (a,d) Schematic of the three-Dirac-fermion model of the nonmagnetic case (a), and the AFM case (d), where the Dirac states ($D_1$ and $D_2$) correspond to the top and bottom surface states of the surface layer, and $D_3$ comes from the top surface states of the below bulk. $\Delta_{12}$ is the intralayer coupling between $D_1$ and $D_2$, while $\Delta_{23}$ is the interlayer coupling between $D_2$ and $D_3$.  (b,c,e,f) Bands at $\Gamma$ as a function of $\Delta_{23}/\Delta_{12}$, corresponding to increasing the interlayer vdW gap, for the nonmagnetic case (b) and the AFM case hosting an AFM surface layer (c) and an FM surface layer with $\Delta_{12}>h$ (e) and with $\Delta_{12}<h$ (f). The $h$ represents the Zeeman coupling strength.}
\label{fig2}
\end{figure*}

\section{II. Effective model analysis} 
The surface states of the nonmagnetic/magnetic TIs under an expansion of the interlayer vdW gap can be captured by a simple low-energy effective model composed of three helical Dirac fermions. As illustrated in Fig.~\ref{fig2}(a), two of the Dirac fermions, labelled as $D_1$ and $D_2$, respectively, come from top and bottom surface states of the surface layer. The third Dirac fermion, labelled $D_3$, comes from the top surface state of the below bulk. Normally, $D_1$, $D_2$, and $D_3$ are local Dirac states with a typical spread ($1\sim2$ nm). With magnetic moments along the out-of-plane direction (the $z$ direction), each Dirac fermion can be described by the low-energy Hamiltonian 
\begin{equation}
H_{D_{i}}=sv( k_x\sigma_y-k_y\sigma_x)+h_{i}\sigma_z,
\end{equation}
where $i=1,2,3$, $v$ is the Fermi velocity, $s=+1 (-1)$ denotes the helicity of $D_1$ and $D_3$ ($D_2$), the Pauli matrices act on the spin subspace, and $h_i$ indicates the Zeeman coupling. We further consider two couplings due to finite wavefunction overlaps: $\Delta_{12}$ between $D_1$ and $D_2$, $\Delta_{23}$ between $D_2$ and $D_3$. Note that $\Delta_{12}$ depends on the thickness of the surface layer, while $\Delta_{23}$ relys on the topmost vdW gap. In the ordered basis of ($|D_{1} \uparrow \rangle$, $|D_{1} \downarrow\rangle$, $|D_{2} \uparrow\rangle$, $|D_{2} \downarrow\rangle$, $|D_{3} \uparrow\rangle$, $|D_{3} \downarrow\rangle$), the total Hamiltonian of the coupled three-Dirac-fermion model is given by
\begin{equation}
H=
\left(
\begin{array}{ccc}
  H_{D_{1}}& \Delta_{12} \sigma_{0}&0\\
\Delta_{12}  \sigma_{0}& H_{D_{2}} & \Delta_{23} \sigma_{0}\\
0& \Delta_{23}\sigma_{0}&H_{D_{3}}
\end{array}
\right),
\end{equation} 
where $\sigma_0$ is $2\times2$ identity matrix in the spin subspace.

We first consider nonmagnetic TIs to see how surface states change with an expansion of the topmost vdW gap ($d$). With increasing $d$ from the pristine bulk value ($d=d_0$), $\Delta_{23}$ gradually decreases to zero, while $\Delta_{12}$ remains unchanged. Therefore, we treat $\Delta_{12}$ as the energy unit hereinafter. Without the Zeeman term, the nonmagnetic three-Dirac-fermion model exhibits two linear bands with dispersions $\pm vk$, where $k=\sqrt{(k_x^2+k_y^2)}$. Regardless of $\Delta_{23}/\Delta_{12}$, the two linear bands always meet at $\Gamma$, forming a gapless Dirac cone, as reflected from the doubly degenerate energy level at shown in Fig.~\ref{fig2}(b). However, the location of the gapless Dirac cone gradually changes from the top of the surface layer to the top of the below bulk [see Fig.~\ref{fig2}(a)]. The persistent gapless Dirac cone in the nonmagnetic case can be understood from the destructive interference behavior in a three-level system.

When A-type AFM order is present, each of the three Dirac fermions becomes gapped due to a mass term induced by the Zeeman coupling. Interestingly, the competition between the Zeeman coupling $h$ and the intralayer coupling $\Delta_{12}$ leads to distinct results. We first consider the surface layer as one FM block with $h_1=h_2=-h_3=h$ (where $h$ denotes the Zeeman coupling strength assumed to be positive). Intriguingly, if the condition $\Delta_{12}>h$ is satisfied, for example, for a sufficiently thin surface layer, there is always an energy-gap-closing-and-reopening process with a gapless Dirac-cone surface state, when the interlayer coupling $\Delta_{23}$ is tuned by modulating the vdW gap, as shown in Fig.~\ref{fig2}(e). In contrast, if $\Delta_{12}<h$, the energy gap remains open, as shown in Fig.~\ref{fig2}(f). We will show that the gapless surface state is topologically protected by the Chern number transition of the three-Dirac-fermion system from the weakly coupled limit of $\Delta_{23}\rightarrow0$ to the strongly coupled limit of $\Delta_{23}\gg\Delta_{12}$.

In the weakly coupled limit of $\Delta_{23}\rightarrow0$, only $D_1$ and $D_2$ are coupled through $\Delta_{12}$, while $D_3$ is nearly isolated with the Chern number $C_3=\textrm{sgn} (h_3)/2=-1/2$. For the coupled $D_1$ and $D_2$, if $\Delta_{12}>h$ ($\Delta_{12}<h$), its Chern number is $C_{12}=0$ [$C_{12}=\mathrm{sgn} (h_{1})=1$]~\cite{Yu2010quantized}, and correspondingly, the total Chern number is $C=C_{12}+C_3=-1/2$ ($C=1/2$). In the strongly coupled limit with $\Delta_{23}\gg\Delta_{12}$, and $\Delta_{23}\gg h$, $D_1$ becomes nearly isolated, and only $D_2$ and $D_3$ are coupled. Because of the $PT$ symmetry combining the spatial-inversion operation ($P$) and the time-reversal operation ($T$) [see the supplementary material (SM)~\cite{SM} for details], the coupled $D_2$ and $D_3$ give a zero Chern number. Therefore, the total Chern number in the strongly coupled limit is equivalent to that of $D_1$, given by $C_1=\textrm{sgn} (h_1)/2=1/2$. It follows that if $\Delta_{12}>h$ is satisfied, the total Chern number of the three-Dirac-fermion system changes by $|\Delta C|=1$ from the weakly coupled limit to the strongly coupled limit in the vdW gap expansion process, ensuring the existence of a topological transition shown in Fig.~\ref{fig2}(e). Differently, for $\Delta_{12}<h$, the total Chern number remains unchanged and the energy gap of surface states stays open [see Fig.~\ref{fig2}(f)]. Therefore, we can conclude that the gapless surface state is topologically protected and arises from the competition between the Zeeman coupling and the Dirac fermion couplings.

Secondly, we consider an AFM surface layer with $h_1=-h_2=h_3=h$. In this case, there is no gapless transition with the interlayer vdW gap expansion, shown in Fig.~\ref{fig2}(c). Instead, the energy gap of the surface state remains an almost constant magnitude $2h$, irrespective of $\Delta_{23}/\Delta_{12}$. Moreover, the total Chern number remains unchanged as $C=1/2$, which can be obtained through similar arguments as above by taking two limits of $\Delta_{23}\ll\Delta_{12}$ and $\Delta_{23}\gg\Delta_{12}$ into account. 

Furthermore, it is worth mentioning that the three-Dirac-fermion model can also be used to describe the FM TIs, for example, by setting $h_1=h_2=h_3=h$. However, in contrast to the A-type AFM TIs, in the FM case, a topological transition with a gapless surface state appears only when $\Delta_{12}<h$ is satisfied (see SM for more details~\cite{SM}).

\begin{figure*}[htbp]
\includegraphics[width=5.8in]{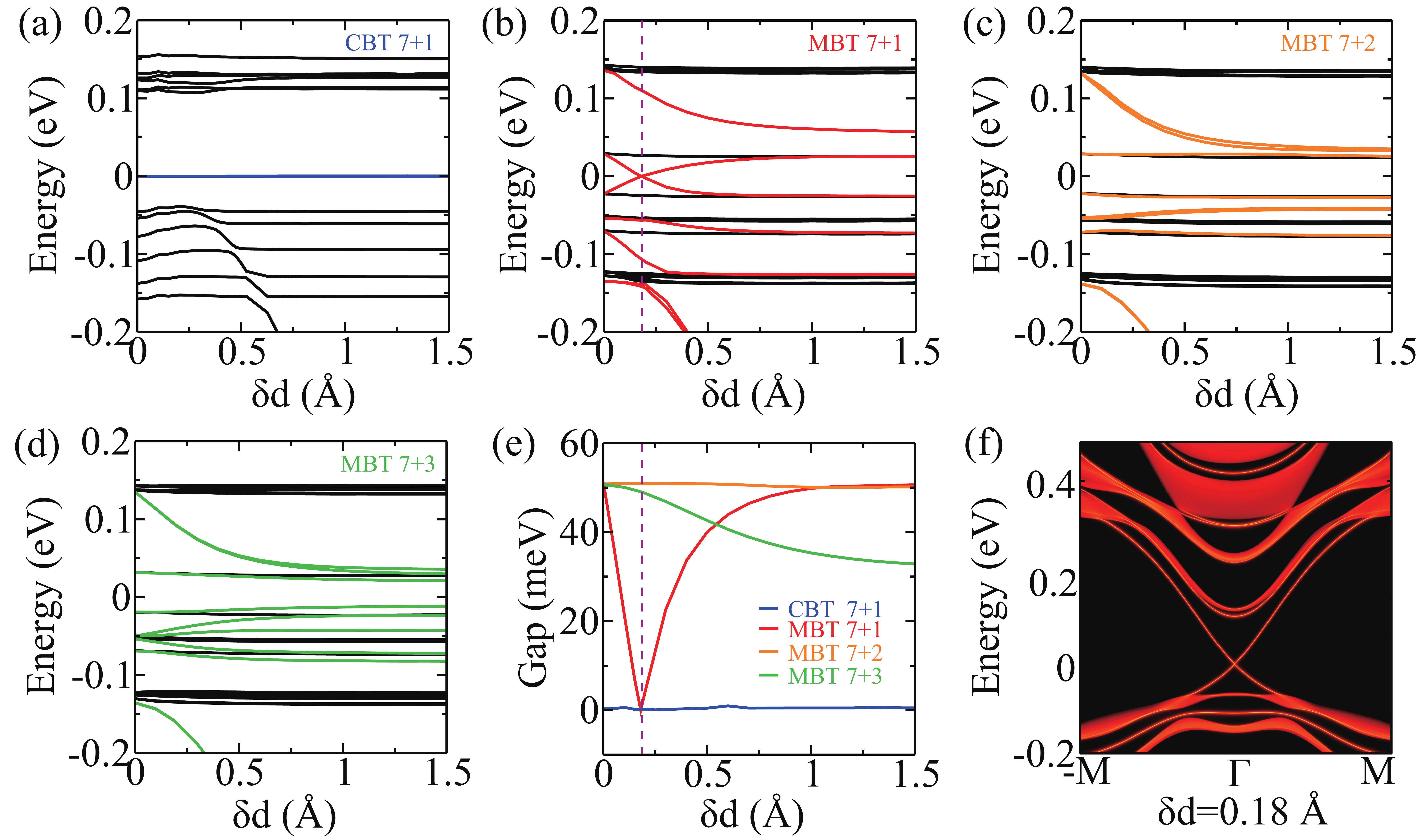}
\caption{AFM TI MnBi$_2$Te$_4$ and nonmagnetic TI CaBi$_2$Te$_4$ with an expansion of the vdW gap. (a-d) Bands at $\Gamma$ for  ($7+1$)-SL CaBi$_2$Te$_4$ (a), ($7+1$)-SL MnBi$_2$Te$_4$ (b),  ($7+2$)-SL MnBi$_2$Te$_4$ (c), and ($7+3$)-SL MnBi$_2$Te$_4$ (d). The bands on the surface layer and the top SL of the below bulk are highlighted in blue, red, orange, and green, respectively. (e) The corresponding band gaps are presented. A gapless transition of the band gap arises at the vdW gap expansion $\delta d=0.18\ \mathrm{\AA}$ for ($7+1$)-SL MnBi$_2$Te$_4$, with the corresponding surface LDOS (f), where gapless Dirac-cone states can be seen.}
\label{fig3}
\end{figure*}

\section{III. Material realization} 
As a concrete example of the three-Dirac-fermion model, in what follows we study the layered AFM TI MnBi$_2$Te$_4$. It is composed of FM SLs that are coupled to each other through the vdW force, and it exhibits an $A$-type AFM order in the magnetic ground state, with out-of-plane FM coupling within each SL and AFM coupling between neighboring SLs~\cite{Zhang2019mbt}. However, the existence of unavoidable Mn$_\textrm{Bi}$ and Bi$_\textrm{Mn}$ antisite defects and Mn vacancies might lead to significant changes of the vdW gap~\cite{Shikin2020nature,Yan2019prm, Zeugner2019chemical,Liang2020prb}. The mechanical cleavage and  exfoliation processes of the sample preparation, and the symmetry suddenly breaking on the surface may also result in a small expansion of the topmost interlayer vdW gap~\cite{Shikin2020nature}. Now, based on first-principles calculations, we investigate the expansion effects of the topmost vdW gap of MnBi$_2$Te$_4$.

\begin{figure*}[htbp]
\begin{center}
\includegraphics[width=5.8in]{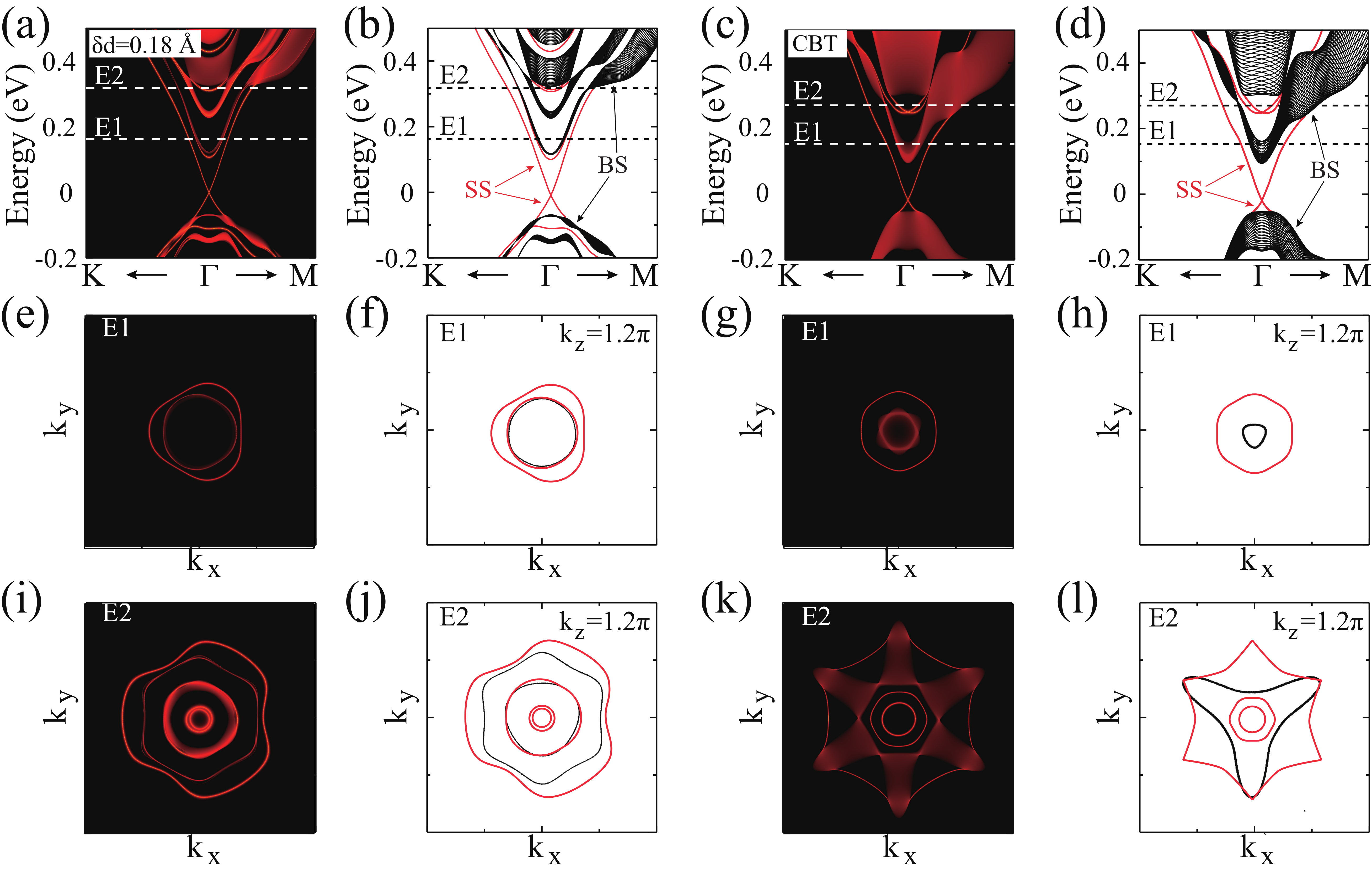}
\end{center}
\caption{Surface LDOSs and Fermi surfaces (FSs) of AFM MnBi$_2$Te$_4$ and nonmagnetic CaBi$_2$Te$_4$. The surface LDOS of MnBi$_2$Te$_4$ (a,b) and FSs at energy levels E1 (e,f) and E2 (i,j) with the topmost vdW gap expansion $\delta d=0.18\ \mathrm{\AA}$. The surface states are gapless in (a,b). Surface FSs show the three-fold rotation symmetry due to the broken TRS, while bulk FSs (e.g. $k_z=1.2\pi$) interestingly show a six-fold-like rotation symmetry, because of the weak $k_z$ dependence of bulk bands due to the $PT$ symmetry \cite{li2019-prb}. The surface LDOS (c,d) and FSs at E1 (g,h) and E2 (k,l) of CaBi$_2$Te$_4$. The surface FSs show a six-fold rotation symmetry, while bulk FSs (e.g. $k_z=1.2\pi$) clearly show a three-fold rotation symmetry. Moreover, notably, there are a set of Rashba-type surface states at E2 due to breaking the inversion symmetry on the surface.}
\label{fig4}
\end{figure*}

We start from the nonmagnetic case of CaBi$_2$Te$_4$ where Mn atoms in MnBi$_2$Te$_4$ are replaced with Ca atoms. For our calculations, we take a ($7+1$)-SL CaBi$_2$Te$_4$ thin film with gradually expanding the topmost vdW gap between the first and second SL. The evolution of energy bands at $\Gamma$ with increasing $\delta d=d-d_0$ is presented in Fig.~\ref{fig3}(a), where the bands locating at the first and second SLs are highlighted in blue. The blue bands at the Fermi level are doubly degenerate, indicating the persistent existence of the gapless Dirac-cone surface states in the first and second SLs, which is well consistent with the model analysis in Fig.~\ref{fig2}(b).

We now investigate the AFM TI MnBi$_2$Te$_4$. We take ($7+1$)-SL and ($7+3$)-SL MnBi$_2$Te$_4$ films for our calculations. Here, 1 SL in ($7+1$)-SL film and 3 SLs in ($7+3$)-SL film are effectively considered as the surface layer, while the remaining 7 SLs are considered as the below bulk. We make these choices based on the following considerations. For the thickness of MnBi$_2$Te$_4$ thin film smaller (greater)  than 3 SLs, the coupling $\Delta_{12}$ between the two Dirac-cone surface states ($D_1$ and $D_2$) is expected to be greater (smaller) than the Zeeman coupling strength $h$, thus leading to a Chern number of 0 ($1$) for the 1-SL (3-SL) MnBi$_2$Te$_4$ film. In Figs.~\ref{fig3}(b) and \ref{fig3}(d), we plot the energy bands at $\Gamma$ with increasing the vdW gap expansion $\delta d$ for ($7+1$)-SL and ($7+3$)-SL MnBi$_2$Te$_4$, respectively, where red and green bands denote the bands from the surface layer. As shown in Fig.~\ref{fig3}(b),  the energy gap closes at a small expansion ($\delta d=0.18 \ \mathrm{\AA}$) and then reopens (red lines) in Fig.~\ref{fig3}(e) with increasing $\delta d$. It is worth mentioning that this critical value of $\delta d=0.18\ \mathrm{\AA}$ is stable and not sensitive with the different functionals of first-principles calculations (see SM~\cite{SM}). In Fig.~\ref{fig3}(f), we explicitly plot the surface local density of states (LDOS) at the critical value of  $\delta d=0.18\ \mathrm{\AA}$, confirming the emergence of gapless Dirac-cone surface state.  In contrast, for the case of the ($7+3$)-SL MnBi$_2$Te$_4$ film, the band gap decreases with increasing $\delta d$, as shown by green lines in Figs.~\ref{fig3}(d) and \ref{fig3}(e), but it tends to saturate for large $\delta d$ and never closes. We also present corresponding results for ($7+2$)-SL MnBi$_2$Te$_4$ in Figs.~\ref{fig3}(c) and \ref{fig3}(e) (orange lines), where the energy gap of the surface states remains almost unchanged throughout the expansion process.  Therefore,  our first-principles calculations for the ($7+1$)-SL, ($7+2$)-SL, and ($7+3$)-SL MnBi$_2$Te$_4$ films are consistent with the predictions of the three-Dirac-fermion model, corresponding to Figs.~\ref{fig2}(e), \ref{fig2}(c) and \ref{fig2}(f), respectively.

\section{IV. Discussion and conclusion} 
For the surface states of MnBi$_2$Te$_4$, there was a problem of experimental and theoretical incompatibility. Unexpected gapless surface states were observed by many ARPES experiments~\cite{Chen2019prx, Li2019prx, Hao2019prx}, contrary to gapped surface states predicted by theories~\cite{Zhang2019mbt,liu2021magnetic}. Very encouragingly, the mechanism of the vdW gap expansion, which we proposed,  can solve this problem. Taking a small expansion of the topmost vdW gap (e.g., $\delta d=0.18 \ \mathrm{\AA}$) for MnBi$_2$Te$_4$, the gapless Dirac-cone surface states of MnBi$_2$Te$_4$ have been obtained by using the present three-Dirac-fermion approach and first-principles calculations. 

A previously proposed explanation for the gapless Dirac-cone surface states is the reconstruction of the surface magnetization, but it will break the three-fold rotation symmetry, resulting in the loss of three-fold rotation symmetry of the surface states~\cite{Hao2019prx,yang2022qpi}. In contrast, under the present mechanism of the vdW gap expansion, the gapless surface states retain the original three-fold rotation symmetry. In Fig.~\ref{fig4}, we plot the calculated result for the surface LDOS and Fermi surfaces (FSs) at two selected energy levels for MnBi$_2$Te$_4$ with $\delta d=0.18 \ \mathrm{\AA}$ (first and second columns). One can see that the surface states are gapless and their FSs preserve the three-fold rotation symmetry but break the six-fold rotation symmetry due to the broken TRS. We also calculate the surface states and FSs of CaBi$_2$Te$_4$ to exclude the possibility of gapless surface states induced from nonmagnetic TIs, as shown in Fig.~\ref{fig4} (third and fourth columns). Though the surface states of CaBi$_2$Te$_4$ are gapless,  they preserve the six-fold rotation symmetry, which is essentially different from the three-fold rotation symmetry in AFM TI MnBi$_2$Te$_4$. It is worth mentioning that the gapped surface states were observed in the recent point contact tunneling spectroscopy on MnBi$_2$Te$_4$~\cite{Ji2021detection}, indicating that a moderate pressure on the surface can reduce an expansion of the vdW gap to obtain gapped surface states. Therefore, it is expected that a moderate pressure will provide a promising way to realize the QAHE with a large band gap in MnBi$_2$Te$_4$ films.

Furthermore, we have applied the three-Dirac-fermion model to other vdW magnetic TIs in the MnBi$_2$Te$_4$ family, such as MnBi$_4$Te$_7$, MnBi$_6$Te$_{10}$, and MnSb$_4$Te$_7$ (see SM~\cite{SM}). It is found that the gapless surface states universally exist in all these materials under the expansion of the topmost vdW gap, which is not only confirmed by our first-principles calculations but also consistent with most experimental observations~\cite{Wu2020Distinct, Vidal2021orbital, Hu2020, Hu2020van, Ma2020}. 

\section{Acknowledgement}
This work is supported by National Key Projects for Research and Development of China (Grant No.2021YFA1400400, No.2017YFA0303203 and No. 2022YFA1403602), the Fundamental Research Funds for the Central Universities (Grant No. 020414380185), Natural Science Foundation of Jiangsu Province (No. BK20200007), the Natural Science Foundation of China (No. 12074181, No. 12104217, No. 11834006, and No. 12174182) and the Fok Ying-Tong Education Foundation of China (Grant No. 161006). D. Wang is supported by the program A/B for Outstanding PhD candidate of Nanjing University. \\ D. W and H. W contributed equally to this work.

\bibliography{ref}

\end{document}


\title{Supplemental Materials for ``Three-Dirac-fermion approach to unexpected universal gapless surface states in van der Waals magnetic topological insulators''}

\author{Dinghui Wang$^{1}$, Huaiqiang Wang$^{1\ast}$, Dingyu Xing$^{1,2}$, and Haijun Zhang$^{1,2\ast}$}

\affiliation{
 $^1$ National Laboratory of Solid State Microstructures, School of Physics, Nanjing University, Nanjing 210093, China\\
$^2$ Collaborative Innovation Center of Advanced Microstructures, Nanjing University, Nanjing 210093, China\\
}

\email{zhanghj@nju.edu.cn; hqwang@nju.edu.cn; }
\maketitle

\tableofcontents
\section{Emergent PT symmetry in coupled two-Dirac-fermion model}
Here, we show the existence of the combined symmetry of inversion $P$ and time-reversal $T$ in a coupled two-Dirac-fermion system, where the two Dirac fermions have opposite T-breaking mass terms (induced by opposite magnetic moments) and opposite helicities. For the three-Dirac-fermion model with $h_1=h_2=-h_3=h$ in the main text,  when $\Delta_{12}=0$, $D_1$ is isolated from $D_2$ and $D_3$, and the coupled $D_2$ and $D_3$ subsystem can be exactly described by the above model. In the ordered bases of ($|D_{2\uparrow} \rangle$,$|D_{2\downarrow} \rangle$,$|D_{3\uparrow} \rangle$,$|D_{3\downarrow} \rangle$), the four-band model Hamiltonian is given by 

\begin{equation}
H(\mathbf{k})=
\left(
\begin{array}{cccc}
   h&   iv k^{-}& \Delta_{23}&0\\
   -iv k^{+}&   -h&0&  \Delta_{23}\\
     \Delta_{23}&   0&-h&-iv k^{-}\\
  0&    \Delta_{23}&iv k^{+}&-h
\end{array}
\right)=-\tau_z(k_x\sigma_y-k_y\sigma_x-h\sigma_z)+\Delta_{23}\tau_x,\tag{S1}
\end{equation} 
where $k^{\pm}=k_x\pm ik_y$, $v$ is the Fermi velocity, $h$ represents the Zeeman coupling strength, and the Pauli matrices $\tau_i$ and $\sigma_i$ act in the orbital and spin subspaces. The inversion and time-reversal operators are given by $\tau_x$ and $i\sigma_y K$ ($K$ mean complex conjugate), respectively. The combined $PT$ operator is then simply obtained as $i\tau_x\sigma_y K$. It is straightforward to show that the above Hamiltonian satisfies the $PT$ symmetry as

\begin{equation}
PT H(\mathbf{k})(PT)^{-1}=H(\mathbf{k}).\tag{S2}
\end{equation}

Since the antiunitary $PT$ operator satisfies $(PT)^2=-1$, Kramers degeneracy is ensured at every momentum, leading to doubly degenerate band structures with vanishing total Berry curvatures and thus zero Chern number for the occupied bands.

\begin{figure}
 \centering
  \includegraphics[width=5.4in]{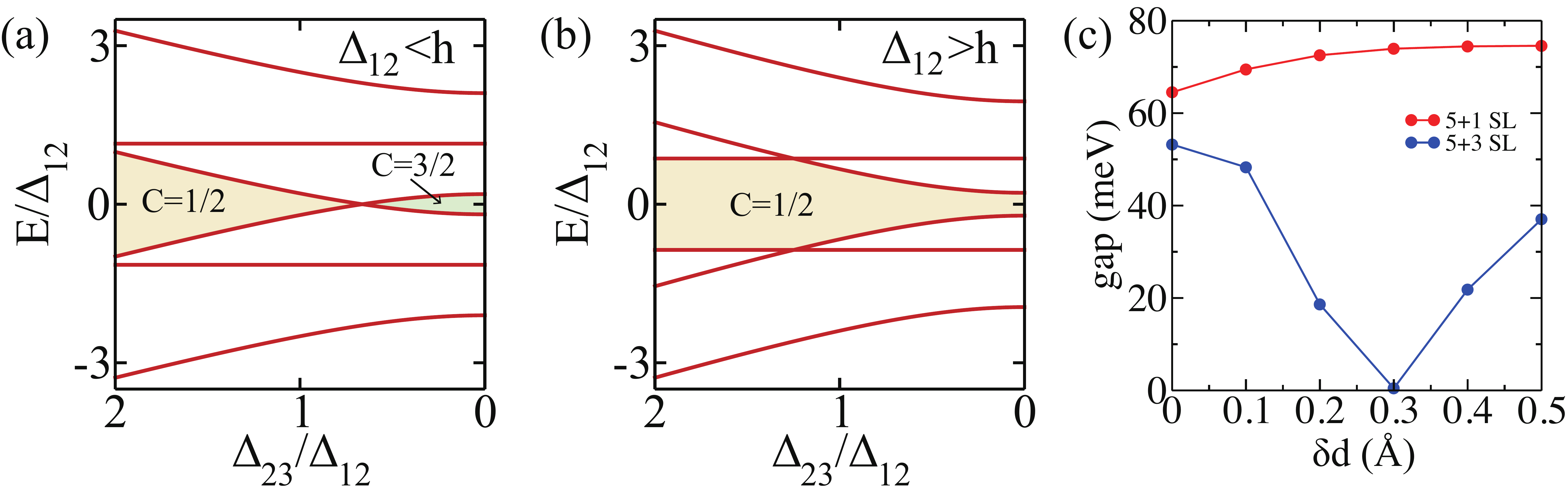}\\
  \caption{Bands at the $\Gamma$ point of the coupled three-Dirac model with FM moments and (a) $\Delta_{12}<h$, (b) $\Delta_{12}>h$ in the expansion process of the van der Waals gap between surface and layers, where $\Delta_{23}/\Delta_{12}$ gradually decreases to zero in the surface-detached limit. The gap closing-and-reopening process appears only when $\Delta_{12}<h$, which is accompanied by a unit change of the Chern number ($C$). In the numerical calculation, $h/\Delta_{12}$ is chosen to be $1.2$ and $0.8$ in (a) and (b), respectively. (c) Surface gap evolution from first-principles calculation of a (5+1)-SL (red lines) and (5+3)-SL (blue lines) FM MnBi$_2$Te$_4$ with out-of-plane FM moments.}\label{figs1}
\end{figure}

\section{Ferromagnetic case by three-Dirac-fermion model}
Here, we present the detailed analysis of the ferromagnetic (FM) counterpart of three-Dirac-fermion model, where the gapless transition point appears only when $\Delta_{12}<h$ is satisfied. For the coupled three-Dirac-fermion model with FM moments, i.e., $h_1=h_2=h_3=h$ ($h>0$ is assumed), the Hamiltonian is given by
\begin{equation}
H=
\left(
\begin{array}{cccccc}
  h&-iv k^{-}&  \Delta_{12}&   0&0&0\\
ivk^{+}&-h&   0&    \Delta_{12}&0&0\\
\Delta_{12}&0&   h&   iv k^{-}& \Delta_{23}&0\\
0& \Delta_{12}&   -iv k^{+}&   -h&0&  \Delta_{23}\\
0&0&     \Delta_{23}&   0&h&-iv k^{-}\\
0&0&   0&    \Delta_{23}&iv k^{+}&-h
\end{array}
\right). \tag{S3}
\end{equation} 
The energy spectrum can be easily obtained as
\begin{equation}
E(\mathbf{k})=\pm\sqrt{h^2+k^2},\pm\sqrt{k^2+(\sqrt{\Delta_{12}^2+\Delta_{23}^2}-h)^2},\pm\sqrt{k^2+(\sqrt{\Delta_{12}^2+\Delta_{23}^2}+h)^2}.
\end{equation}
Obviously, the energy gap closes at the $\Gamma$ point when $\sqrt{\Delta_{12}^2+\Delta_{23}^2}=h$ is satisfied. 

In the detaching process of the surface system with the van der Waals gap distance between surface and bulk layers changing from $d_0$ to $\infty$, as shown in Fig. 2(a) in the main text, $\Delta_{12}$ remains unchanged, while $\Delta_{23}/\Delta_{12}$ gradually decreases from a large value to zero. It follows that, when $\Delta_{12}<h$, there exists a gap closing-and-reopening process at a critical value of $\Delta_{23}/\Delta_{12}$, as shown in Fig. S1(a). This is accompanied by a change of the total Chern number from 1/2 to 3/2. In contrast, when $\Delta_{12}>h$,  the gap always keeps open, as shown in Fig. S1(b), and the Chern number keeps unchanged as 1/2. Note that the condition for the mergence of gap closing in the FM case is opposite to that in the antiferromagnetic (AFM) case in the main text. 

To confirm the above analysis based on the three-Dirac-fermion model, we have carried out first-principles calculations of (5+1)-SL and (5+3)-SL FM MnBi$_2$Te$_4$ with out-of-plane magnetic moments, where the topmost 1 SL and 3SL are gradually detached from the the 5SL bulk. The evolutions of the surface gap size in the expansion process of the vdW gap distance between the detached layer and the bulk are shown in Fig. S1(c), where a gapless point emerges for the (5+3)-SL case with $\Delta_{12}<h$, while the gap remains open for the (5+1)-SL case with $\Delta_{12}>h$. These results are consistent with the theoretical analysis of the FM case by the three-Dirac fermion model.

\section{Surface LDOS and Fermi contours by three-Dirac-fermion model}
In this section, to further show the functionality of the three-Dirac model, we will use it to calculate the surface local density of states (LDOS) and Fermi contours in the vdW gap expansion process for both the nonmagnetic case and AFM case exhibiting a gapless transition point, corresponding to the (7+1)-SL nonmagnetic CaBi$_2$Te$_4$ and AFM MnBi$_2$Te$_4$ films in the main text.

\begin{figure}
 \centering
  \includegraphics[width=5.4in]{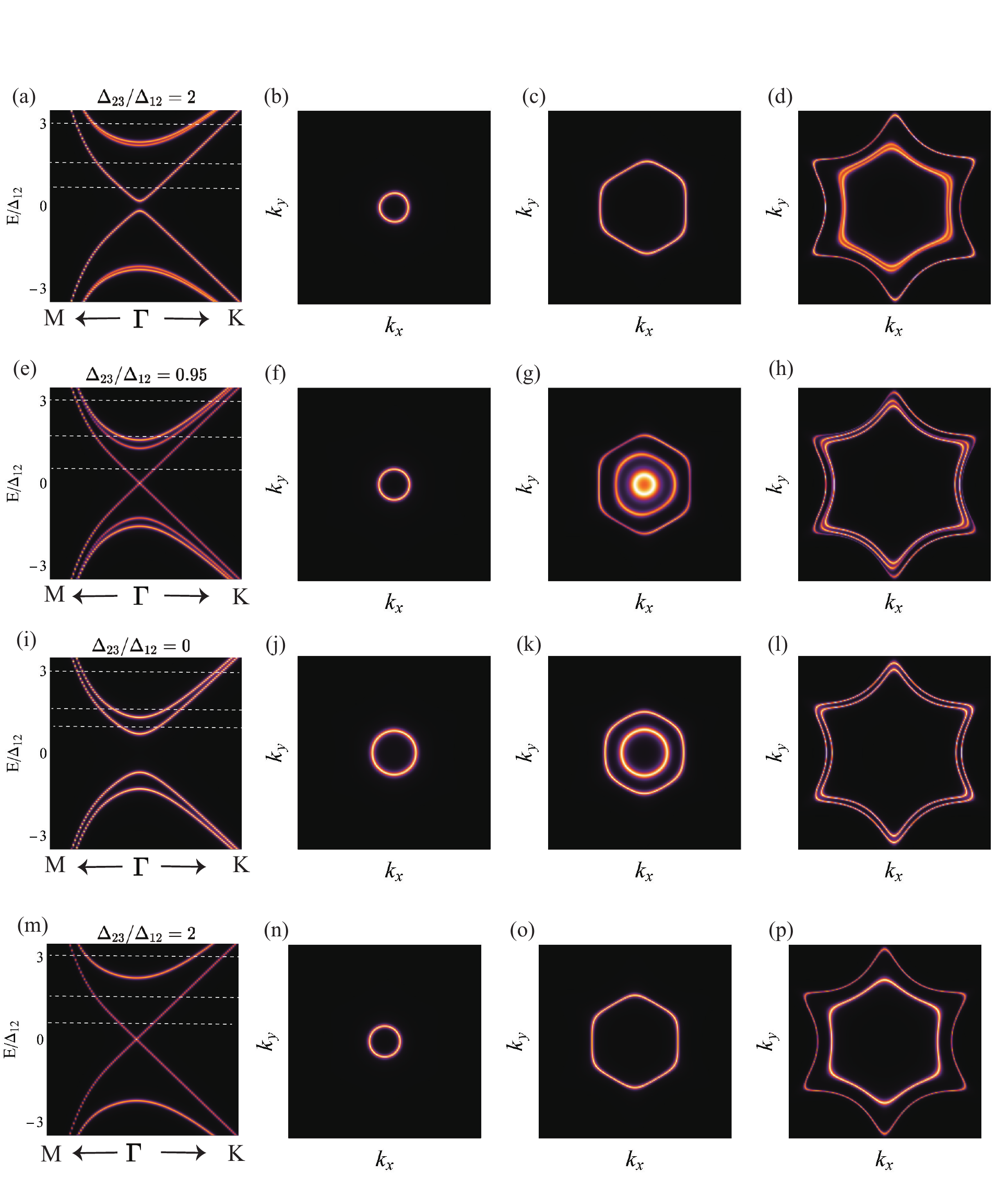}\\
  \caption{Surface LDOS (first column) and corresponding Fermi contours in ascending order of the three energy cuts. The upper three rows correspond to the three stages in the interlayer vdW gap expansion process of a layered AFM topological insulator, which presents the gapless transition point in the expansion process. (First row) the unexpanded case with a representative value of $\Delta_{23}/\Delta_{12}=2$. (Second row) at the critical gap expansion value of $\Delta_{23}/\Delta_{12}\approx 0.95$.  (Third row) the surface-detached limit with $\Delta_{23}/\Delta_{12}=0$. The last row describes the nonmagnetic case with a representative $\Delta_{23}/\Delta_{12}=2$. In the numerical calculations, the Zeeman field strength is set as $h/\Delta_{12}=0.3$, and the coefficient $\lambda$ of the third-order terms in Eq. (S4) is chosen as $\lambda_1=-\lambda_2=\lambda_3$ with $\lambda/\Delta_{12}=0.2$.}\label{figs2}
\end{figure}

To reveal the trigonal or hexagonal structures of the surface states, we need to include symmetry-allowed higher-order corrections up to third-order terms of $k$ in the Hamiltonian of Eq. (2) in the main text.  A nonmagnetic Bi$_2$Te$_3$-type surface state satisfies the following three symmetries, namely, the time-reversal symmetry $T$, three-fold rotation symmetry $C_{3z}$, and mirror symmetry $M_x$. In the presence of the out-of-plane magnetic order, both $T$ and $M_x$ are broken, whereas the $C_{3z}$ symmetry and combined time-reversal and mirror symmetries $M_xT$ are preserved. Through the method of invariant in the $k\cdot p$ theory, all symmetry allowed terms up to third order in $k$ can be obtained (see, e.g., Ref.~\cite{naselli2022magnetic}). For simplicity, we consider the following symmetry allowed third-order correction for each of the three Dirac fermions
\begin{equation}
H'_{D_i}=\lambda_i(k_{+}^3+k_{-}^3)\sigma_{z},\tag{S4}
\end{equation}
where $k_{\pm}=k_x\pm ik_y$. It is worth mentioning that this third-order correction term for Dirac surface states of TIs has also been derived in Ref. \cite{Liu2010Model}. The full Hamiltonian $H'$ of the coupled three-Dirac-fermion system is then obtained by combining the Hamiltonian in Eq. (2) of the main text with the above terms.

To calculate the LDOS and Fermi contours of the gradually detached surface composed of the two Dirac fermions, $D_1$ and $D_2$, we resort to the Green's function method, and the surface LDOS is given by 
\begin{equation}
\rho(\omega)=-\frac{1}{\pi}\textrm{Im} G^{r}_{11,\sigma}(\omega)-\frac{1}{\pi}\textrm{Im} G^{r}_{22,\sigma}(\omega),\tag{S5}
\end{equation} 
where the retarded Green's function $G^{r}(\omega)$ is given by  $G^{r}(\omega)=(\omega+i\eta-H')^{-1}$. 

In Fig. S2, we present the results of surface LDOS (first column) and three representative Fermi contours in ascending order of the three energy cuts in the LDOS for the nonmagnetic layered topological insulators (TIs) (last row) and three stages in the interlayer vdW gap expansion process of layered AFM TIs, namely, without vdW expansion (first row with a representative $\Delta_{23}/\Delta_{12}=2$), at the critical gap expansion value exhibiting the gapless Dirac cone (second row with $\Delta_{23}/\Delta_{12}\approx 0.95$ for $h/\Delta_{12}=0.3$), and, in the surface-detached limit with $\Delta_{23}/\Delta_{12}=0$ (third row). From the third column [Figs. S2(c, g, k)], it can be clearly seen that, with increasing vdW gap expansion, the Fermi contours gradually change from trigonal shape to hexagonal shape, which results from the gradually recovered inversion symmetry of the detached surface system. It should be emphasized that the results predicted by the effective three-Dirac-fermion model are in qualitative consistence with the first-principles results in Fig. 4 in the main text. 

We remark that, in real materials, there also exist Rashba-like bands on the surface, e.g., due to the broken inversion symmetry on the surface. However, since we focus on the Dirac surface states originating from nontrivial bulk topology, which suffice to explain the trigonal or hexagonal warping effect in magnetic TIs, we did not include these Rashba surface bands in the above theoretical discussion by three-Dirac-fermion model.

\section{Results from different vdW functionals}
To further validate the emergence of a gapless point in the expansion process of the topmost interlayer vdW gap of a $(7+1)$-SL MnBi$_2$Te$_4$, we have chosen different vdW functionals to plot the evolution of the surface gap, as shown in Fig. S3. Obviously, the gapless point stably emerges in all cases.

\begin{figure}
 \centering
  \includegraphics[width=3in]{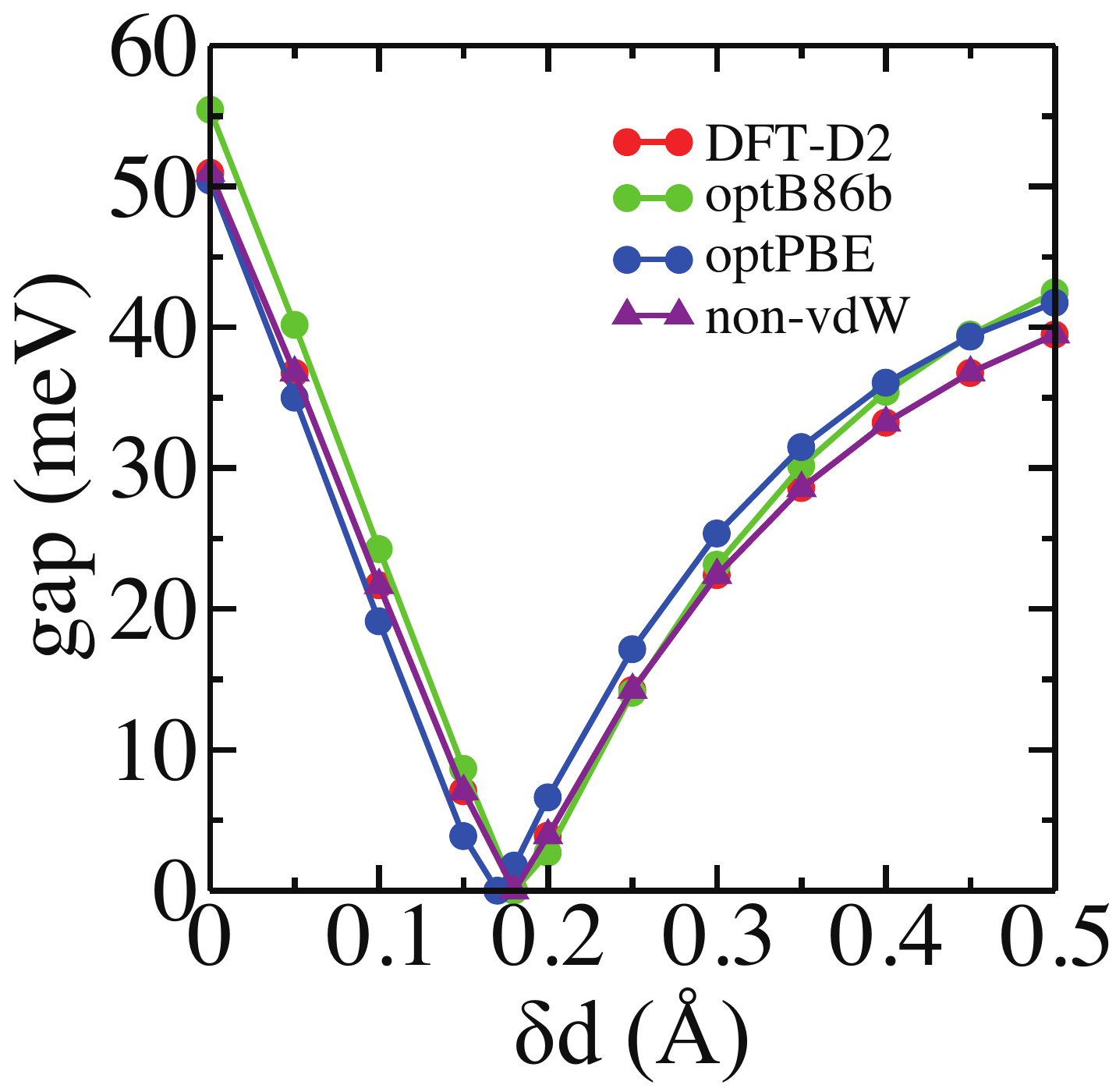}\\
  \caption{The evolution of the surface band gap in the expansion process of the topmost interlayer vdW gap of a (7+1)-SL MnBi$_2$Te$_4$ with different vdW functionals. All curves exhibit a gapless point around $0.18\ \AA$ in the expansion process.}\label{figs3}
\end{figure}

\section{Results for other vdW magnetic topological insulators}
In this section, to further show the functionality and universality of the proposed three-Dirac-fermion model, we apply it to some other representative examples of vdW magnetic TIs in the  MnBi$_2$Te$_4$ family, namely, MnBi$_4$Te$_7$, MnBi$_6$Te$_{10}$, MnSb$_2$Te$_4$ and MnSb$_4$Te$_7$. Since the basic physics is similar, we here focus on MnBi$_4$Te$_7$ and present results from both first-principles calculations and three-Dirac-fermion model analysis, which are consistent with each other. Whereas, we only present the first-principles results for the rest three materials.
  
\subsection{MnBi$_4$Te$_7$}
As shown in Fig. S4(a), MnBi$_4$Te$_7$ consists of one SL MnBi$_2$Te$_4$ and one quintuple-layer (QL) Bi$_2$Te$_3$ as the building block. Similar to MnBi$_2$Te$_4$, the magnetic ground state features out-of-plane FM couplings within each SL, and AFM couplings between adjacent SLs. MnBi$_4$Te$_7$ is proposed to be a nontrivial $Z_2$ AFM TI, which is expected to exhibit gapped Dirac surface states for boundaries along the out-of-plane (z-) direction. However, experimental observations show that~\cite{Wu2020Distinct, Vidal2021orbital, Hu2020, hu2020van}, depending on the surface terminations, there exist two types of topological surface states for MnBi$_4$Te$_7$, namely, gapless (gapped) surface states for SL (QL) termination. The gapped surface states for the QL termination case have been explained to result from the hybridization between surface and bulk bands. However, the gapless surface states for the SL termination still remain elusive. 

Here, we will take advantage of the three-Dirac-fermion model to show that  in the vdW expansion process of the topmost vdW gap distance, a gapless transition point in the surface band structure can emerge only for the SL-termination case, whereas the surface bands always remain gapped for the QL termination. It should be emphasized that our results are in consistence with the experimental observations, and can provide further insight into the properties of the surface state in MnBi$_4$Te$_7$.

\subsubsection{SL termination case}
We first consider the SL termination case, where the three-Dirac-fermion models consists of two Dirac surface states from the topmost SL and the other from the top surface state from the QL below, as shown in Fig. S4(a). The model Hamiltonian is then given by
\begin{equation}
H=
\left(
\begin{array}{cccccc}
  h_{\textrm{SL}}&-iv k^{-}&  \Delta_{12}&   0&0&0\\
ivk^{+}&-h_{\textrm{SL}}&   0&    \Delta_{12}&0&0\\
\Delta_{12}&0&   h_{\textrm{SL}}&   iv k^{-}& \Delta_{23}&0\\
0& \Delta_{12}&   -iv k^{+}&   -h_{\textrm{SL}}&0&  \Delta_{23}\\
0&0&     \Delta_{23}&   0&h_{\textrm{QL}}&-iv k^{-}\\
0&0&   0&    \Delta_{23}&iv k^{+}&-h_{\textrm{QL}}
\end{array}
\right). \tag{S6}
\end{equation} 
Here the Zeeman term $h_{\textrm{SL}}$ results from the magnetic moments within the SL, while the Zeeman term $h_{\textrm{QL}}$ comes from the FM proximity effect on the QL from neighboring SLs. Note that $h_{\textrm{SL}}$ can be treated as a constant, whereas $h_{\textrm{QL}}$ depends on the interlayer distance between the QL and neighboring SLs. Since a QL has two neighboring SLs, $h_{\textrm{QL}}$ in Eq. (S6) consists of two contributions, $h_{\textrm{QL}}^{I}$ and $h_{\textrm{QL}}^{II}$, from the upper and lower SLs, respectively, which depend on the distance between the QL and neighboring SLs. For simplicity and without loss of generality, in the unexpanded case, we assume the proximity-induced Zeeman term for the top surface state of the QL to be $h_{\textrm{QL}}^{I}=0.5h_{\textrm{SL}}$ ($h_{\textrm{QL}}^{II}=-0.3h_{\textrm{SL}}$) from the upper (lower) SL. In the expansion process of the topmost interlayer vdW gap between the QL and the SL above, $h_{\textrm{QL}}^{II}$ obviously keeps unchanged, while $h_{\textrm{QL}}^{I}$ gradually decreases until it vanishes in the decoupled limit, which can be assumed to be proportional to the distance-dependent coupling $\Delta_{23}$ between the QL and the SL. As a result, $h_{\textrm{QL}}$ is then given by 
\begin{equation}
h_{\textrm{QL}}=h_{\textrm{QL}}^{I}+h_{\textrm{QL}}^{II}=0.5h_{\textrm{SL}}(\Delta_{23}/\Delta_{23}^{0})-0.3h_{\textrm{SL}},
\end{equation}
where $\Delta_{23}^0$ denote the coupling $\Delta_{23}$ in the unexpanded case with $d=d_0$.

With the above assumptions, the evolution of the SL-terminated surface band structure at the $\Gamma$ point in the topmost vdW gap expansion process can be obtained, as shown in Fig. S4(b), where $\Delta_{12}$ is chosen as the energy unit and the other parameters are representatively set to be $h_{\textrm{SL}}/\Delta_{12}=0.5$ and $\Delta_{23}^{0}/\Delta_{12}=2$. A gap closing-and-reopening process is discernable with a gapless point separating two phases with $C=1/2$ and $C=-1/2$, respectively. Similar to the discussion in the main text, the $C=1/2$ phase can be understood from the $\Delta_{23}\gg\Delta_{12}$ limit, where the total Chern number is reduced to that of the gapped Dirac fermion $D_1$ on top of the SL with $C=(1/2)\ \mathrm{sgn}(h_{\textrm{SL}})=1/2$. Whereas $C=-1/2$ comes from the other limit of $\Delta_{23}=0$, where the total Chern number is given as the sum of the isolated SL and the top surface state D3 of the QL. For the isolated SL with $h_{\textrm{SL}}<\Delta_{12}$, its Chern number equals zero, so the total Chern number is reduced to that of D3 with $C=(1/2)\ \mathrm{sgn}(h_{\textrm{QL}}^{II})=(1/2)\ \mathrm{sgn}(-h_{\textrm{SL}})=-1/2$. Consequently, such a gapless point is topologically protected and thus inevitable for the SL termination case in the vdW gap expansion process.

As a further evidence, in Fig. S4(c), we plot the surface band gap evolution from first-principles calculations of a AFM MnBi$_4$Te$_7$ with 4 (SL+QL) layers and SL termination,  where the topmost 1 SL is gradually detached from the bulk. The emergence of the gap closing point in Fig. S4(c) is consistent with the above three-Dirac-fermion model analysis.

\begin{figure}
 \centering
  \includegraphics[width=6in]{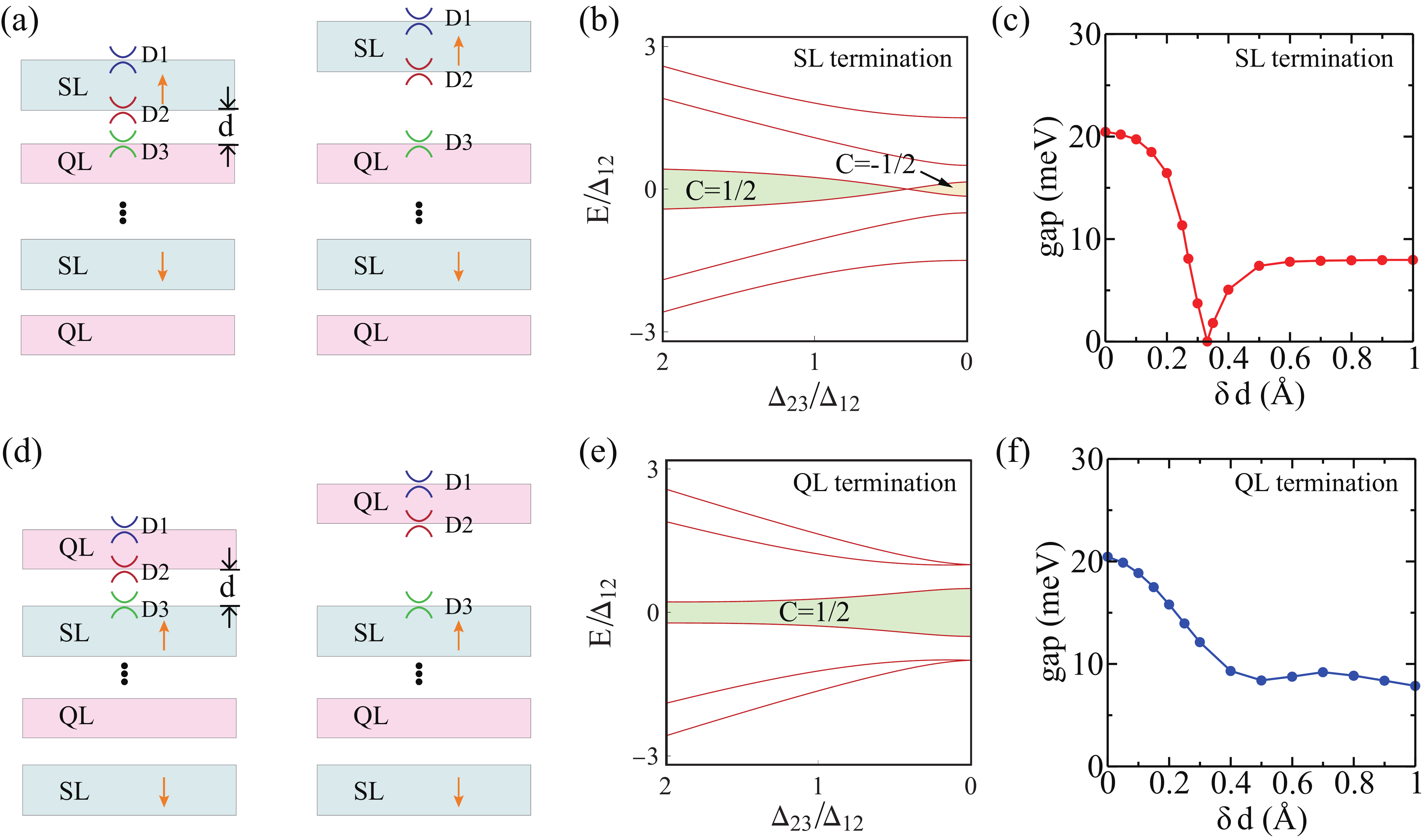}\\
  \caption{Schematic of the three-Dirac model of the SL termination case (a), and QL termination case (d) of AFM MnBi$_4$Te$_7$.  Bands at $\Gamma$ as a function of $\Delta_{23}/\Delta_{12}$, corresponding to increasing the interlayer vdW gap from the three-Dirac-fermion model for the SL termination case (b) and the QL termination case (e).   First-principles results of the surface gap evolution in the topmost vdW expansion process with increasing $\delta d$ for a 4-(SL+QL) MnBi$_4$Te$_7$ with the SL termination case (c) and the QL termination case (f). Gapless point only appears for the SL termination case.}\label{figs4}
\end{figure}

\subsubsection{QL termination case}
For the QL termination case, as shown in Fig. S4(e), the three-Dirac-fermion model is composed of two Dirac fermions D1 and D2 from the two surface states of the QL, and D3 from the top surface state of the SL. The Hamiltonian can then be written as  

\begin{equation}
H=
\left(
\begin{array}{cccccc}
  h_{\textrm{QL}}^{t}&-iv k^{-}&  \Delta_{12}&   0&0&0\\
ivk^{+}&-h_{\textrm{QL}}^t&   0&    \Delta_{12}&0&0\\
\Delta_{12}&0&   h_{\textrm{QL}}^b&   iv k^{-}& \Delta_{23}&0\\
0& \Delta_{12}&   -iv k^{+}&   -h_{\textrm{QL}}^b&0&  \Delta_{23}\\
0&0&     \Delta_{23}&   0&h_{\textrm{SL}}&-iv k^{-}\\
0&0&   0&    \Delta_{23}&iv k^{+}&-h_{\textrm{SL}}
\end{array}
\right). \tag{S7}
\end{equation} 
Here, $h_{\textrm{QL}}^t$ and $h_{\textrm{QL}}^b$ represent the proximity-effect-induced Zeeman term of the top and bottom surface states of the topmost QL from the SL below. Similar to the discussion above, both $h_{\textrm{QL}}^t$ and $h_{\textrm{QL}}^b$ depend on the interlayer distance between the topmost QL and the below SL, and they are also proportional to the interlayer coupling $\Delta_{23}$. For simplicity, $h_{\textrm{QL}}^t$ and $h_{\textrm{QL}}^b$ are reasonably assumed to be  $0.3h_{\textrm{SL}}(\Delta_{23}/\Delta_{23}^{0})$ and $0.5h_{\textrm{SL}}(\Delta_{23}/\Delta_{23}^{0})$, respectively. The evolution of the band structure at $\Gamma$ in the expansion process of the topmost interlayer vdW gap distance can then be obtained, shown in Fig. S4(e). Obviously, the surface band gap remains open and the Chern number keeps unchanged as $1/2$. This result can also be understood by inspecting the two limits of $\Delta_{23}/\Delta_{12}\rightarrow\infty$ and $\Delta_{23}/\Delta_{12}\rightarrow 0$. In the $\Delta_{23}/\Delta_{12}\rightarrow\infty$ ($\Delta_{23}/\Delta_{12}\rightarrow 0$) limit, the total Chern number is reduced to that of D1 (D3) on top of the QL (SL), which is simply given as $C=(1/2)\ \mathrm{sgn} (0.3h_{\textrm{SL}})=1/2$ [$C=(1/2)\ \mathrm{sgn} (h_{\textrm{SL}})=1/2$]. It follows that the band gap should remain open with no gapless transition point. This is also further confirmed by the first-principles calculation of the surface gap evolution of a 4 (SL+QL) MnBi$_4$Te$_7$ with QL termination, as shown in Fig. S4(f).

\begin{figure}
 \centering
  \includegraphics[width=4in]{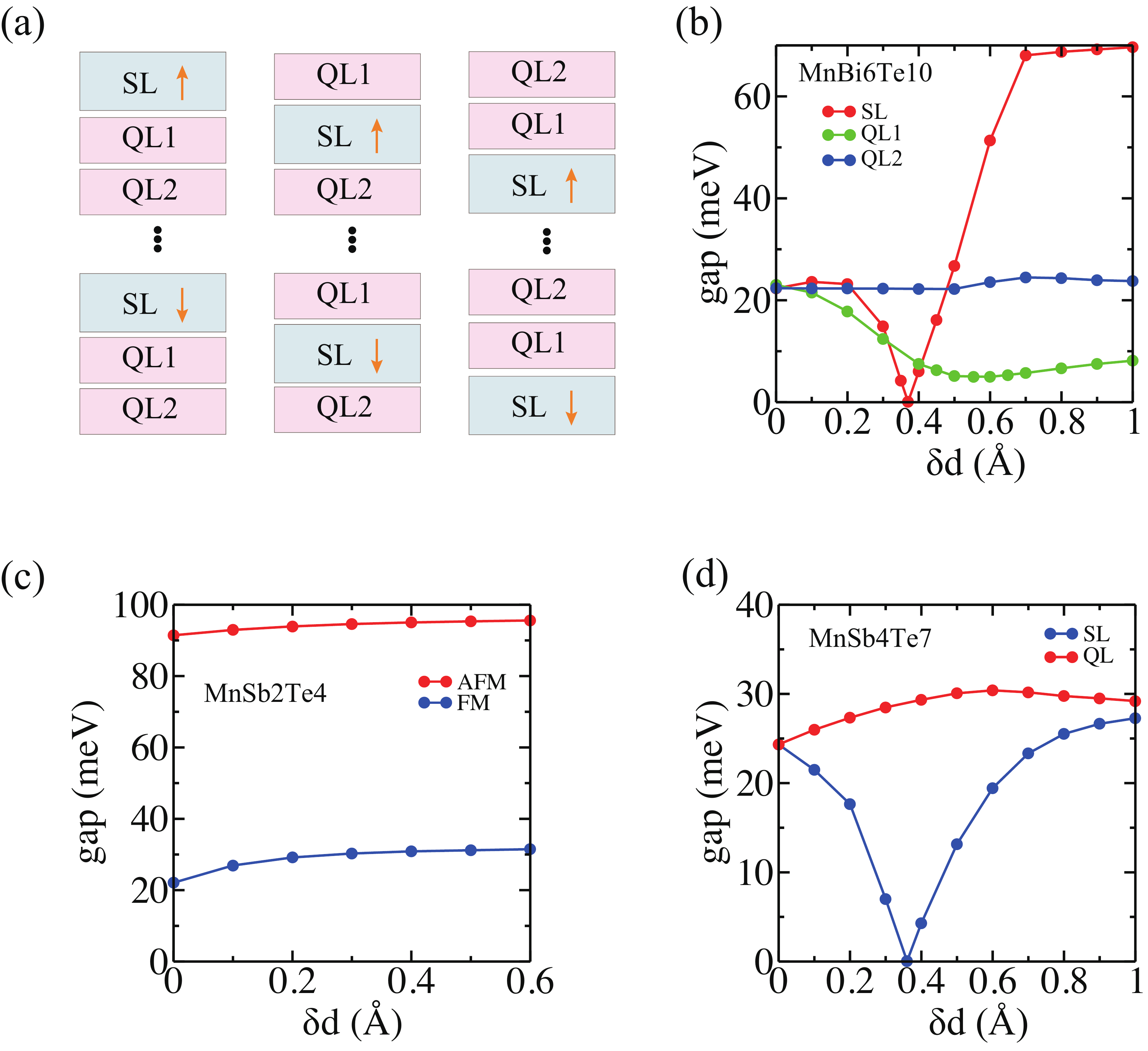}\\
  \caption{(a) Schematic of MnBi$_6$Te$_{10}$ with three different kinds of surface terminations. Surface band gap evolutions in the expansion process of the topmost interlayer vdW gap from first-principles calculations for 3-(SL+QL+QL) AFM MnBi$_6$Te$_{10}$ with the three surface terminations (b), (5+1)-SL MnSb$_2$Te$_4$ with both FM and AFM configurations (c), and 4-(SL+QL) AFM MnSb$_4$Te$_7$ with SL and QL terminations, respectively (d).}\label{figs5}
\end{figure}

\subsection{MnBi$_6$Te$_{10}$, MnSb$_2$Te$_4$ and MnSb$_4$Te$_7$}
Considering the similar underlying physics concerning the three-Dirac-fermion model, here, we only present the results of the surface band gap evolutions in the topmost interlayer vdW gap expansion process from first-principles calculations for MnBi$_6$Te$_{10}$, MnSb$_2$Te$_4$ and MnSb$_4$Te$_7$.

As schematically shown Fig. S5(a), MnBi$_6$Te$_{10}$ has three types of surface terminations, namely, SL-, QL1-, QL2- terminations, respectively. In Fig. S5(b), we choose 3-(SL+QL+QL) AFM MnBi$_6$Te$_{10}$ with different surface terminations to plot the surface band gap as a function of the expansion $\delta d$ of the topmost interlayer vdW gap. It can be seen that similar to MnBi$_4$Te$_7$, only the SL termination of MnBi$_6$Te$_{10}$ exhibit a gapless point, while the gap remains for both QL1 and QL2 terminations.

Figure S5(c) presents the results of the band gap evolution for both the AFM and FM cases of a (5+1)-SL MnSb$_2$Te$_4$. Since MnSb$_2$Te$_4$ is topologically trivial without any Dirac surface states, the band gap persists in the topmost vdW gap expansion process for both cases.

In fig. S5(d), we plot the evolutions of the surface band gap in the topmost vdW gap expansion process of a 4-(QL+SL) AFM MnSb$_4$Te$_7$ films with SL and QL terminations, respectively. Because MnSb$_2$Te$_4$ is topologically nontrivial, analogous to MnBi$_4$Te$_7$, the gapless point appears only in the SL termination case.

\section{Wavefunction profile of topological surface states}
To further support the validity of the three-Dirac-fermion model in the description of layered TIs with vdW gap modulation, we plot the wavefunction profile of the topological surface state in the vdW gap expansion process. The wavefunction of the Dirac state at the $\Gamma$ point is simply assumed as $\psi_i\left(z\right)\propto e^{-\lambda_i\left|z-z_i\right|}$ ($1/\lambda_i$ characterizes the material-dependent decay length and $\lambda_i=1$ is chosen for simplicity in numerical calculations), with $i=1, 2$, and $3$ corresponding to $D_1$, $D_2$, and $D_3$ Dirac states, respectively, and $z_i$ denotes the position of the peak of each Dirac state. When there exist finite overlaps between wavefunctions of these Dirac states, e.g., when the surface layer is sufficiently thin (e.g., a few SLs for MnBi$_2$Te$_4$), finite couplings will be induced between them. This underlies the coupling terms of $\Delta_{12}$ between $D_1$ and $D_2$, and $\Delta_{23}$ between $D_2$ and $D_3$ in the construction of the coupled three-Dirac-fermion model, shown in Fig. S6(a). From this coupling picture, it is obvious that the final topological surface state should be formed by the hybridization of all the three Dirac states.

Here, we take the case of $\Delta_{12}>h$ as an example, corresponding to Fig. 2(e) in the main text, where the coupling between $D_1$ and $D_2$ is significantly larger than the magnetic Zeeman coupling strength. We plot the wavefunction profile of the final topological surface state (red lines) for three representative cases, namely, $\Delta_{23}/\Delta_{12}=2$ [Fig. S6(b)], $\Delta_{23}/\Delta_{12}=0.92$ [Fig. S6(c)], and $\Delta_{23}/\Delta_{12}=0.2$ [Fig. S6(d)], at different expansion distances of the interlayer vdW gap between the surface layer (shaded in pink) and blow bulk (shaded in blue), corresponding to the pristine case with strongly-coupled surface and bulk systems, the critical case near the gapless point with moderate expansion, and the significantly expanded case with weakly-coupled surface and bulk systems. As expected, in the strongly-coupled (weakly-coupled) case, the wavefunction distribution of the topological surface state locates mainly within the surface layer (around the top of the below bulk). Whereas for the moderate expansion case, the topological surface state exhibit similar weights on the surface layer and the top of the below bulk. The overall feature of the above topological surface state is consistent with previous works on MnBi$_2$Te$_4$.

\begin{figure}
 \centering
  \includegraphics[width=5.5in]{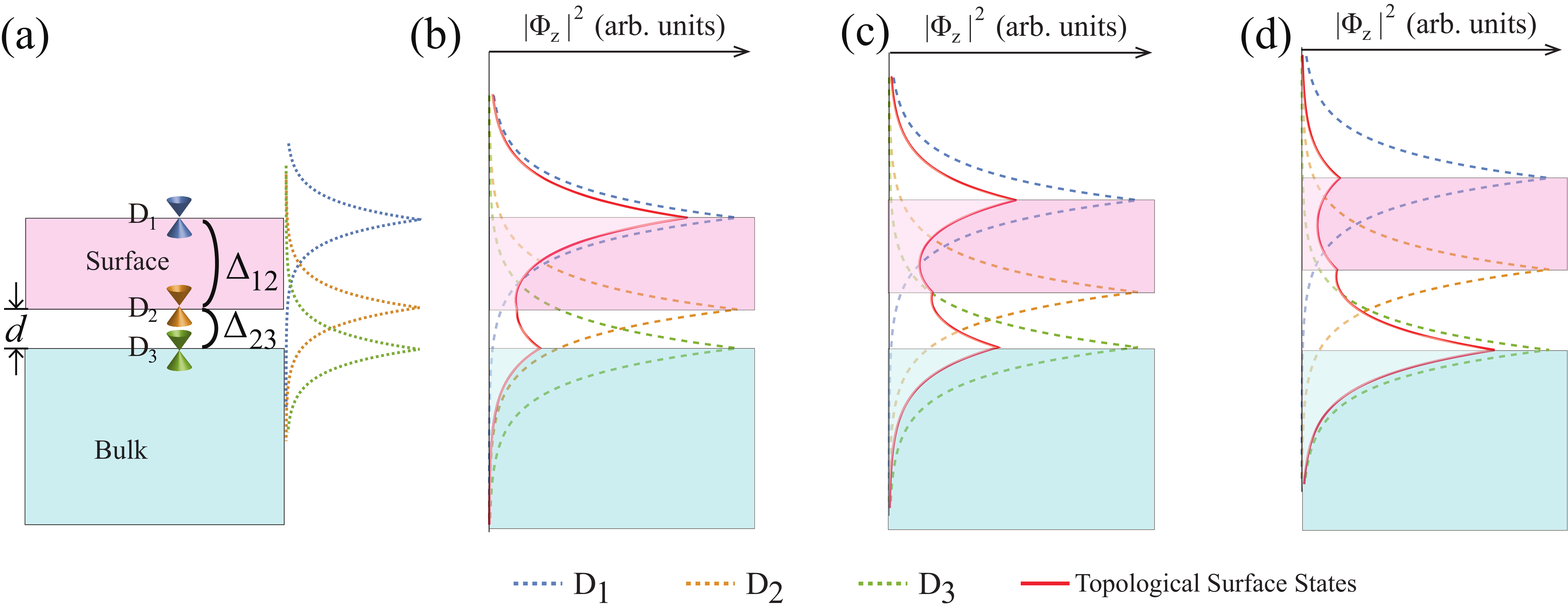}\\
  \caption{For the case of $\Delta_{12}>h$ in the three-Dirac fermion model illustrated in (a), wavefunction profile at $\Gamma$ point of the surface states for three representative cases, $\Delta_{23}/\Delta_{12}=2$ (b), $\Delta_{23}/\Delta_{12}=0.92$ (c), and $\Delta_{23}/\Delta_{12}=0.2$ (d), in the vdW gap expansion process with gradually increased $d$ and decreased $\Delta_{23}/\Delta_{12}$, corresponding to the  the pristine case with strongly coupled surface and bulk systems, the critical case near the gapless point with moderate expansion, and the significantly expanded case with weakly coupled surface and bulk systems. $h=0.4$ is chosen in the numerical calculations.}\label{figs6}
\end{figure}

\section{First-principles calculation details}
First-principles calculations were performed using the Vienna ab initio simulation package (VASP)~\cite{kresse_efficient_1996}. The Perdew-Burke-Ernzerhof (PBE) functional~\cite{perdew_generalized_1996}, within the projector augmented wave (PAW)~\cite{blochl_projector_1994} are used to describe the exchange-correlation potential and energy. Lattice constants are adopted from experimental data, $a=4.334 \text{ \AA}$ and $c=13.637 \text{ \AA}$ for each septuple layer. Then we relax atoms positions. A Hubbard-like $U=5$ eV is used to account for strongly localized Mn 3d orbitals. A kinetic cutoff energy of 410 eV and a 10$\times$10$\times$1 $\Gamma$-centered k points mesh is used in a self-consistent field. The unit cells are stacked along $c$ direction to generate multi-septuple layers films with a $20 \text{ \AA}$ vacuum space for the thick slab system. In all calculations, we keep G-type AFM with a magnetic axis along the $c$ direction. By replacing Mn in MnBi$_2$Te$_4$ with Ca, and then relaxing the atoms positions, we simulate the non-magnetic case in CaBi$_2$Te$_4$. To account for the surface layer vdW expansion effect, we construct the maximally localized Wannier function (MLWF) \cite{marzari_maximally_1997} from first-principles calculations with Mn $d$, Bi $p$, and Te $p$ orbitals. The tight-binding Hamiltonian is divided into surface and bulk parts. The spectral functions and Fermi surface are calculated with the surface Green’s functions of the semi-infinite system. Both spectral functions and the Fermi surface are projected to the upmost septuple layer.

\bibliography{suppleref}